\documentclass[pre,aps,eqsecnum,amsmath,amssymb,showkeys]{revtex4}
\usepackage[sort&compress]{natbib}
\usepackage{bm}
\usepackage{enumitem}
\usepackage{hyperref} 
\usepackage[final]{graphicx}
\usepackage{subfigure}
\usepackage{epstopdf}
\usepackage{color}
\usepackage[]{ulem}

\newcommand{\ADD}[1]{{#1}}

\newcommand{\REM}[1]{{}}

\begin{document}
\title{Spreading of non-motile bacteria on a hard agar plate: Comparison between agent-based and stochastic simulations}
	\author{Navdeep Rana}
	\author{Pushpita Ghosh}\email{pghosh@tifrh.res.in}
	\author{Prasad Perlekar}\email{perlekar@tifrh.res.in}
	\affiliation{\ADD{Tata Institute of Fundamental Research, Centre for Interdisciplinary Sciences, Hyderabad 500107, India}}
	\begin{abstract}
		We study spreading of a non-motile bacteria colony on a hard agar plate by using  agent-based and continuum models. We show that the spreading dynamics depends on the initial nutrient concentration, the motility and the inherent demographic noise. Population fluctuations are inherent in an agent based model whereas, for the continuum model we model them by using a stochastic Langevin equation. We show that the intrinsic population fluctuations coupled with non-linear diffusivity lead to a transition from Diffusion Limited Aggregation (DLA) type morphology to an Eden-like morphology on decreasing the initial nutrient concentration. 
	\end{abstract}
	\maketitle
\section{Introduction}
Pattern formation is perhaps one of the most fascinating aspect in a broad range of natural phenomena \cite{Cross_1993, RevModPhys2013,Kondo1616}. Bacteria in a Petri dish environment  exhibit a large variety of complex spatial patterns ranging from compact circular growth, concentric rings to long branched patterns \cite{wak94,woo95,sha95,ben97,ben98,ver08,kor10,den14}. The colony morphology depends upon various factors such as nutrient concentration, cell motility, growth-proliferation and death dynamics, and other chemical and physical  variables \cite{mit97,har03,shi04,kai07,fau12,kum13,def14,Chiara2015,wu15}. In a classic experiment, Wakita et al. \cite{wak94} obtained the phase-diagram of \textit{Bacillus subtilis} colony morphology as a function of nutrient concentration and solidity of agar medium and identified five basic morphologies: (A) diffusion limited aggregation (DLA), (B) Eden-like, (C) concentric ring-like, (D) homogeneous spreading, and (E) dense branching morphology (DBM).
Similar morphological patterns have also been observed in growing yeast colonies \cite{Sams1997}.
Several studies \cite{ben94,Kawasaki1997,gol98,Ben-Jacob2000,PGhosh2013,Pintu2016,sch16}, since then, have proposed mathematical models to investigate the phase-diagram of Ref. \cite{Kawasaki1997}. 
These models can be broadly classified into two categories: 
\begin{enumerate}[label={(\roman*)}]
	\item Agent based models --- In these models each bacteria is treated as an entity and the collective spatiotemporal behavior largely depends upon the local interactions among them. These interactions can arise from mechanical forces exerted by bacteria as they grow, divide and push each other and spread on a hard substrate. How individual interactions turn out to be significant in forming collective orders, have been explored successfully by using agent based models in some of the earlier studies~\cite{kor11,Farrell2013,PGhosh2015}. Ref.\cite{kor11} utilized an agent-based model of spatial population genetics to explore the role of demographic noise and genetic drifts in bacteria population. Farrell and co-workers ~\cite{Farrell2013} have used  \ADD{an} agent-based model to explore mechanically-driven growth of non-motile rod-like bacteria in an expanding colony which undergoes transitions from circular to branched morphologies with varying nutrient consumption rate or nutrient concentration. Recently, in Ref.\citep{PGhosh2015} mechanical-driven spontaneous phase-segregation of nonmotile, rodshaped bacteria and in presence of self-secreted extracellular polymeric substances in a growing biofilm has been explored using \ADD{an} agent-based model.
	
	\item Reaction diffusion equations --- Here we treat bacteria colony density and nutrient concentration as fields and write continuum equations for them. The bacteria motility and the nutrient spreading is modeled by a diffusive term and the birth and death is modeled with a reaction term. Perhaps the most widely used reaction-diffusion equation is the Fisher equation~\cite{fis37} [Eq.~\eqref{fkpp}] which has been successfully used to model homogeneous spreading (type-D morphology) of bacteria on a soft-agar plate and in a nutrient rich environment. 
	\begin{equation}
	\begin{aligned}
	\partial_t \rho &= D\nabla^2 \rho + \gamma \rho \left(1-\frac{\rho}{Z} \right),
	\label{fkpp}
	\end{aligned}  
	\end{equation}
	where $\rho$ is the bacteria colony density and $Z$ is its carrying capacity. Several studies have incorporated the effect of the nutrient concentration and bacteria motility by coupling Eq.~\eqref{fkpp} with an additional 
	equation for each of these variables to obtain different morphological patterns that were discussed above \cite{wak94,woo95,sha95,ben97,ben97b,ben98,leg04,ver08,kor10,den14,cha16}.  Studies designed to investigate the role of demographic noise use the stochastic variants of Eq.~\eqref{fkpp} \cite{kol37,doe03,kor10,per11a,ben12,pig13}. 
\end{enumerate}

However, how and to what extent nutrient concentration, nutrient diffusivity, growth-proliferation and inherent population fluctuations altogether govern the microbial growth dynamics, morphological trends are yet to be explored in details. In this paper, we explore the morphological spatial dynamics of non motile bacteria growing on a hard agar plate with varying initial nutrient concentration and diffusivity. Our numerical investigation using agent based and continuum models are designed to mimic the experiments of Wakita et al. \cite{wak94} which show a transition from branching to an Eden like pattern. \ADD{Our approach is different than earlier studies \cite{Lacasta1999,ben97,ben97b,ben98,Schwarcz2016} where the bacterial morphological patterns were attributed to substrate properties such as irregularities on the agar substrate \cite{Lacasta1999}, substrate hardness that depends on the agar concentration and local lubrication created by bacteria \cite{Schwarcz2016}, and nutrient concentration.  However, these models ignore the role of population fluctuations that have been shown \cite{Kessler1998,kor10,nes14} to play a crucial rule in determining the growth, competition and cooperation in bacterial colonies under nutrient rich conditions.}

In Section ~\ref{sec:ind} we utilize an agent-based model to investigate the growth dynamics and morphological trends of non motile rod-shaped bacteria growing and spreading by consuming a diffusive nutrient on a hard agar surface.  \ADD{The substrate is assumed to be uniform and frictionless.} This agent-based model automatically takes care of finite-size and particle nature of the organisms.  Unlike in Ref~\cite{Farrell2013}, the nutrient resource is limited and initially kept fixed and uniform in our model in a Petri-dish like set up. We find that growth and morphological dynamics of growing colony depends upon the interplay of local nutrient availability, nutrient diffusivity and mechanical interactions. Colonies growing on a nutrient rich substrate show a rapid growth and a smooth front (type-B) morphology whereas, those growing on a nutrient deficient substrate show slower growth and branched or finger-like structures (type-A) at the front. In contrast to Ref~\cite{Farrell2013}, we also find that nutrient diffusivity can affect bacteria growth dynamics. For fixed resources, reducing nutrient diffusion leads to a slowly growing colony with larger final size whereas high nutrient diffusivity leads to rapidly growing small colonies.

\ADD{Motivated by the results from our agent-based model, in Section \ref{sec:fish} we present a continuum model to study the role of nutrient concentration on spreading of bacteria colony. We assume substrate to be uniform and without inhomogeneities and do not consider substrate-bacteria interaction.}  Our numerical experiments show that population fluctuations and nutrient dependent bacterial diffusivity  destabilize the front and leads to formation of finger-like patterns in nutrient deprived conditions. Similar to the agent based model, we find that increasing initial nutrient concentration leads to a faster growing colony. We show that the front speed follows the mean field predictions. The front structure undergoes a transition from a branching pattern to an Eden pattern  on increasing the initial nutrient condition. We conclude by contrasting the similarities and the differences between the agent-based and the continuum-model.

\section{Agent based model}
\label{sec:ind}
We consider an agent-based model \cite{Farrell2013,PGhosh2015} of nonmotile bacterial cells to study colony growth on a hard agar plate. Individual cells are represented by a growing sphero-cylinder of \ADD{constant diameter ($d_{0} = 1 \mu m$)} and variable length $l$. We consider a two dimensional semi-solid square surface of length $L \equiv 200 \mu$m  (unless otherwise stated in the text), for colony growth. The location of bacterial cell is represented by a two dimensional spatial coordinate ${\bm r} = (x, y)$ and the orientation of its major axis is determined by two unit vectors $( u_x , u_y )$. In our model, the growth of a cell depends on its size and the local concentration of the diffusing nutrient. The initial nutrient concentration is fixed to $C_0$ on all grid points. Bacteria consumes nutrients proportional to it's area and grows which leads to the governing equation for nutrient concentration,
\begin{eqnarray}
\frac{\partial c}{\partial t} = D\left( \frac{\partial^{2} c}{\partial x^2} + \frac{\partial^{2} c}{\partial y^2}\right) - k_c\sum A_{i}f(c(x_i, y_i)) ,
\end{eqnarray} 
where $A_i = \pi r_{0}^{2} + 2r_{0}l_{i}$ is the area of i$^{th}$ individual, $r_0 = d_0/2$ is the radius of end-caps, $l$ is the length of the cell and $x_i, y_i$ are its spatial coordinates. The nutrient is utilized by the microbial cells at a constant rate $k_c f(c)$ per unit biomass density where $f(c)$ is a monotonically increasing dimensionless function. We choose $f(c)=c/(1+c)$, a monod function with half-saturation constant equal to one, \ADD{i.e. concentrations are measured in units of half-saturation constant}. In our model individual bacteria grows along it's major axis as per the relation ${dl_{i}}/{d t} = \phi(A_{i}/\bar{A})f[c(x_i, y_i)]$ where $\phi$ is the constant growth parameter and $\bar{A}= \pi r_{0}^{2} + \frac{3}{2}r_{0}l_{max}$  is the average area \cite{Farrell2013,PGhosh2015}. Once a cell reaches a critical length $l_{max} $, \ADD{it stops growing further and} divides at a rate $k_{div} $ into two independent daughter cells.
\ADD{The orientation each of daughter cell can be different than that of the mother cell because of various environmental factors like slight bending of the cells, elastic forces between cells etc. To achieve this, we give small random kicks to the orientations of daughter cells, after the division. This also prevents the cells from growing in long filament like structures.}
The length of the daughter cells is chosen such that the combined length of the two daughter cells is equal to the length of the mother cell. This criterion fixes the length of the daughter cell to $l_d=(l_{max}-d_0)/2$. \ADD{This represents symmetric division which occurs in most bacteria. However, there are scenarios where asymmetric division can occur in some bacteria~\cite{Rubin}, which we do not consider in the present study.}  

\begin{table}[h!]
		\centering
	\begin{tabular}{|c|c|c|} 
		\hline
		Parameter &  Symbol & Simulations \\ [0.5ex] 
		\hline\hline
		
		Maximum length  & $l_{max}$ & $3.0\mu $m  \\
		
		Diameter of cell & $d_0$ & $1.0 \mu $m   \\
		
		Linear growth rate & $\phi $ & $1.5 \mu m$ $ hr^{-1}$ \\
		
		Cell-division rate & $k_{div}$  & $ 0.1  hr^{-1} $ \\
		
		Elastic modulus of alive cells & $E$ & $3\times10^{5} Pa$\\
		
		Friction coefficient & $\zeta $ & $200 Pa.hr$ \\
		
		Nutrient consumption rate & $ k_c $ & $6.0 hr^{-1}$   \\
		
		Diffusion rate of nutrient & $ D $ & $1 \mu m^{2} hr^{-1}$ \\[1.0ex] 
		\hline
	\end{tabular}
	\caption{Parameters and constants used in the agent-based model}
	\label{table:1}
\end{table}

In our model, individual cells interact directly by mechanical interaction in accordance with the Hertzian theory of elastic contact \cite{Volfson2008} by repulsive forces with neighbouring cells in case of spatial overlap. In a dense colony of nonmotile bacteria, inertial forces can be neglected and we consider only the over-damped dynamics as given by the equation of motion~\citep{Farrell2013}:
\begin{eqnarray}
\dot{\bm r} &=&\frac{1}{\zeta l}{\bm F},\\ 
{\omega} &=& \frac{12}{ \zeta l^3}\tau
\end{eqnarray}
where $\zeta$ is the friction per unit length of cell. Bacterium position and the angular velocity are represented by ${\bm r}$ and $\omega$ respectively. The corresponding linear forces and torques are ${\bm F}$ and $\tau$.
The force between two spherocylinders is approximated by the force between two spheres placed along the major axis of the rods at such positions that their distance is minimal \cite{Farrell2013,PGhosh2015}. If the closest distance of approach between the two nearby spherocylinders is $r$, such that $h=d_0 - r$ is the overlap, then the force magnitude is assumed to be ${F}=Ed_{0}^{1/2}h^{3/2}$, where $E$ parametrizes the strength of the repulsive interaction proportional to the elastic modulus of the cell. $E\longrightarrow \infty $ implies perfectly hard cells but in reality we use a finite value of $E$ (see Table~\ref{table:1}) in our simulations, allowing for some deformation of the cells. In addition to the direct mechanical cell-cell interaction there is a competition for the local nutrient which can be considered as an indirect interaction between microbial cells mediated by the environment. \ADD{All the agent-based simulations are performed in two-dimensional square box with periodic boundary conditions. As initialization, a few number ($N=300$) of bacteria cells of same aspect ratio and random orientations are placed in a 1D inoculation along the line between the two points $(0,L/2)$ and ($L,L/2$), in a narrow strip of about $l_{max}$ in $y$-direction. We use a simple Euler method for the time-evolution of the equations of motion i.e. Eq(2.2) and (2.3) and a central finite-difference scheme to solve the nutrient diffusion in Eq(2.1).}

\begin{figure}[ht!]
	\includegraphics[width=16cm]{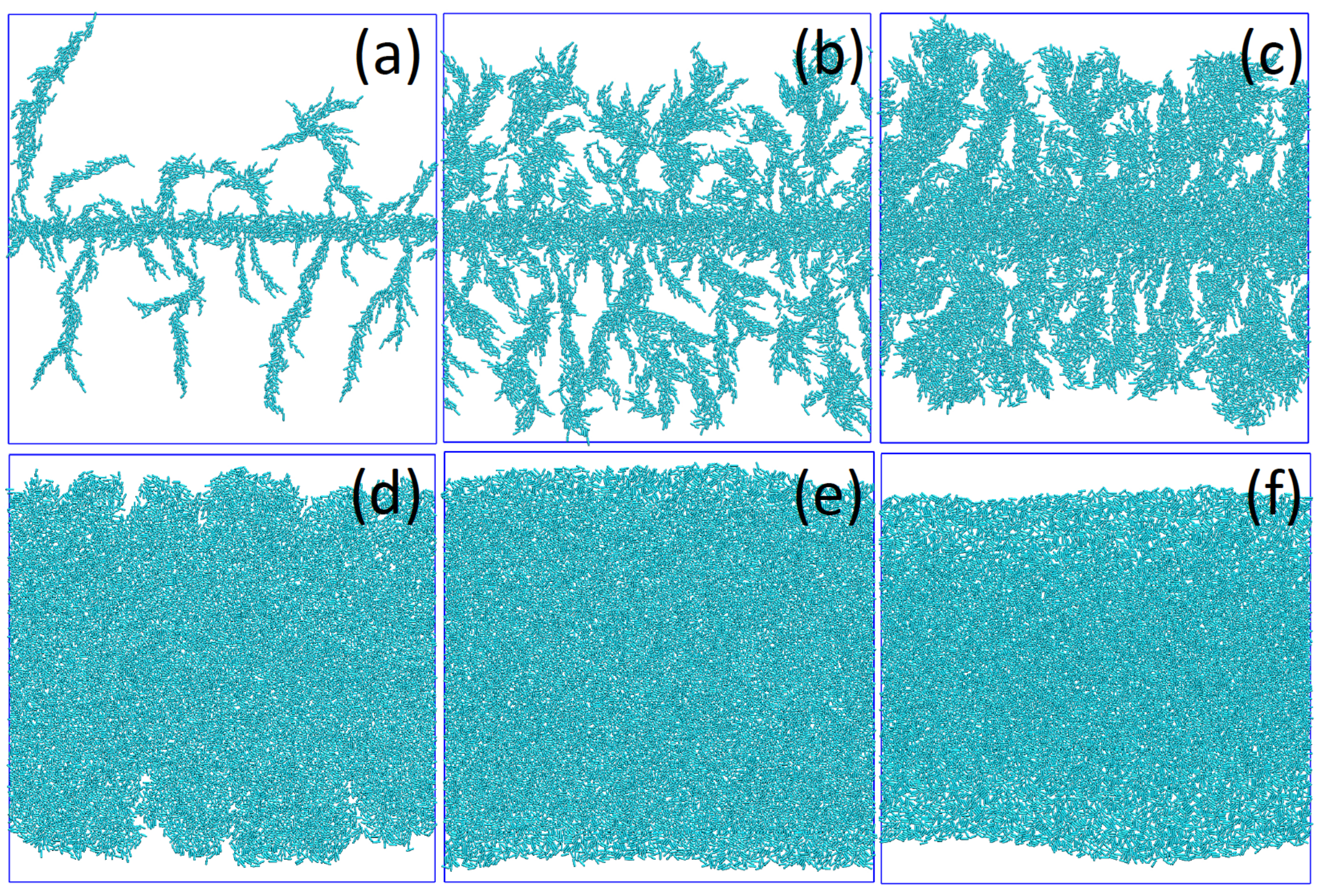} 
	\caption{Cell growth morphologies: Snapshots of well-developed colonies obtained from our agent-based model for different values of initial nutrient concentrations: (a) $C_0$=10, (b) $C_0=50$, (c) $C_0=100$, (d) $C_0=200$, (e)$C_0=300$ and (f) $C_0=400$.  All the other parameters are chosen to be the same as given in the Table-\ref{table:1}.
	}\label{Figure1_varying_nutrient}
\end{figure}
\subsection{Results: Morphology and Speed} 
The major focus of the present study is to understand the role of initial nutrient concentration and nutrient diffusivity on the growth dynamics and morphology of a colony. We begin our study by placing a few number ($\sim 300$) of cells in a 1D inoculation along the line formed by joining the points $(0,L/2)$ and $(L,L/2)$. We vary the initial nutrient concentrations $C_0$  from very low $C_0=10$ to very high  $C_0=400$ values while keeping all the other parameters fixed as given in the Table-\ref{table:1}.  
Fig.~\ref{Figure1_varying_nutrient} demonstrates different morphologies of growing colonies with the variation of $C_0$. We find that for small $C_0=10$, the colony front develops finger-like patterns. As we increase the initial nutrient concentration from $C_0=50$ to $C_0=200$, finger-like patterns are replaced by branched structures. On further increase of initial nutrient concentration to very high values $C_0 > 250$, the rough branched fronts are replaced by smoother colony fronts. Note that for all the cases simulated above, the cells at the front grow by consuming nutrients while the rest of the cells behind the front stop their growth due to complete depletion of nutrient and become frozen.

\begin{figure}[htp]
	\includegraphics[width=0.95\linewidth]{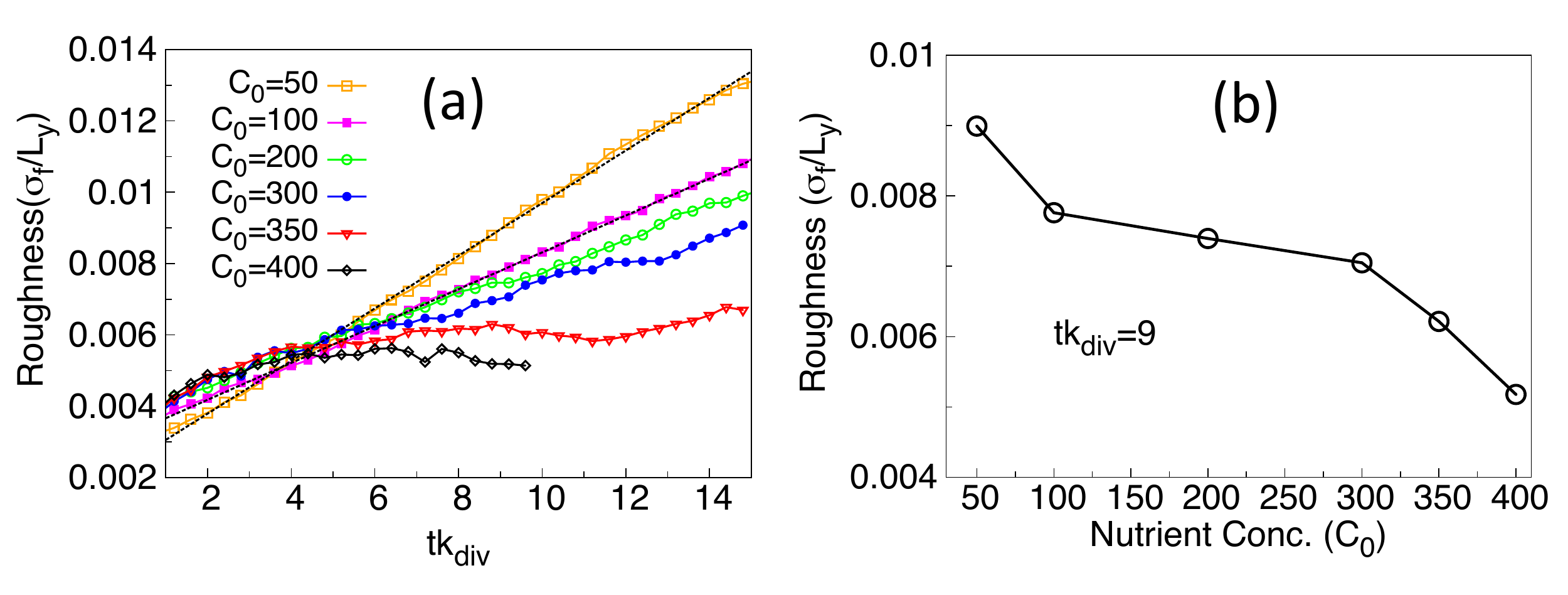} 
	\caption{Roughness of the colony fronts. (a) The normalized standard deviation of the front height(Roughness $\sigma_f/L_y$) where $L_y = 400$, is shown with respect to scaled time for different values of $C_0$s. The black dashed line indicates a linear fit. (b) Roughness of the front is plotted against different values of $C_0$ at $tk_{div}=9$. All the other parameters are chosen to be the same as given in Table-\ref{table:1}.
	}\label{front_roughness}
\end{figure}
\begin{figure}[bth]
\includegraphics[width=.95\linewidth]{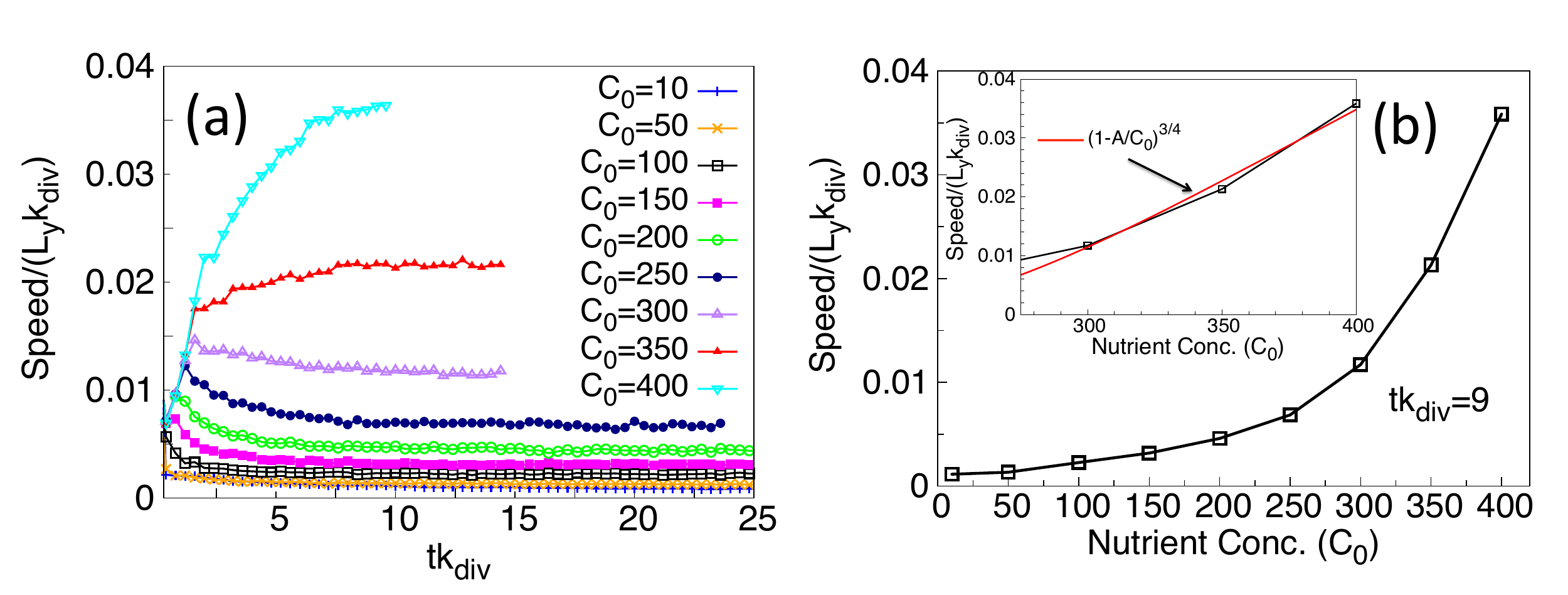} 
\caption{Front speed ($V_f$) of the spreading microbial colonies. (a) For different values of initial nutrient concentrations $C_0$, the speed vs time curves are shown. Speed is rescaled by dividing it by the length of the box in spreading direction($L_y=400$) and rate of cell-division $k_{div}$. Time is rescaled by multiplying with $k_{div}$. (b) \ADD{Plot of the asymptotic spreading speed at $tk_{div}=9$ as a function of initial nutrient concentration. The figure in the inset shows corresponding dependence(red solid line) of speed on $C_0$ at the branching transition regime.} All the other parameters are chosen to be the same as given in Table-1.  
}\label{front_speed}
\end{figure}
To quantify the changes in the growth dynamics and the colony morphology we calculate a roughness parameter $\sigma_f$ which is the ensemble averaged standard deviation of height of the colony front.The front height is determined as follows. We discretize the simulation domain along the X-direction into equal bins of size comparable to the length of daughter cell($l_d$), and find out an individual $i$ whose $x_i$ belongs to the bin and $y_i$ is maximum, then the height of the front at the bin is set to $y_i$. In Fig.~\ref{front_roughness}(a), we plot roughness versus time for different values of $C_0$. For large values of $C_0 \geq 350$, $\sigma_f$ is very small and almost constant in time whereas, it increases in time for $C_0 < 350$ indicating formation of finger or branched structures. \ADD{For smaller values of $C_0 \leq 200$, we find a linear variation of $\sigma_f$ with respect to time.} To further quantify the variation in the front thickness in well-developed colonies, in Fig. \ref{front_roughness}(b), we plot $\sigma_f$ after nine generations (\ADD{$tk_{div}=9$}). We find that the front roughness increases with decreasing $C_0$ in agreement with the morphological trends shown in Fig.~\ref{Figure1_varying_nutrient}.

We now investigate how nutrient limitation influences the speed at which a colony spreads. \ADD{The front speed is calculated as ensemble average of the rate of change of the covered area $A_C$ over the box length ($L_y$) in the spreading direction Y, i.e. speed $V=\frac{1}{L_y}  \langle dA_C/dt \rangle$ where $\langle \rangle$ denote ensemble averaging.} 
The plot in Fig. \ref{front_speed}(a) shows that the asymptotic front speed of the colony increases with increasing $C_0$. For large $C_0$, the initial increase in speed is because of abundance of nutrients at $t=0$ which leads to rapid cell divisions both in the bulk as well as the colony front. The front speed achieves the asymptotic value when the nutrient consumption balances the diffusion. \ADD{In Fig. \ref{front_speed}(b), we plot the asymptotic spreading speed at ninth generation ($tk_{div}=9$) for different values of $C_0$ and observe that colonies on nutrient rich substrate spread faster. We also observe that for $C_0 > C_\star$, $V \sim \alpha [1-C_\star/C_{0}]^{3/4}$ \cite{Farrell2013} where  $\alpha$ is a fitting parameter and $C_\star=192$ is the approximate value of the concentration at which we observe colony morphology transition from branched to uniform.} 
\subsection{Results: Nutrient diffusivity} 
\begin{figure}[htp!]
	\begin{subfigure}	\centering
		\includegraphics[height=0.25\linewidth]{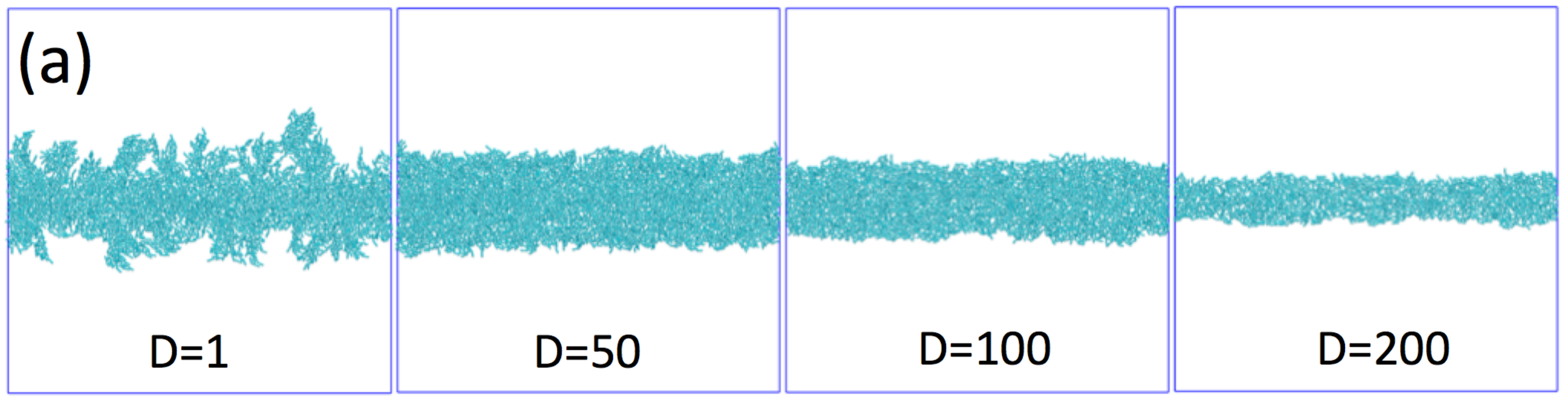}
	\end{subfigure}
	\begin{subfigure}	\centering
		\includegraphics[width=0.475\linewidth]{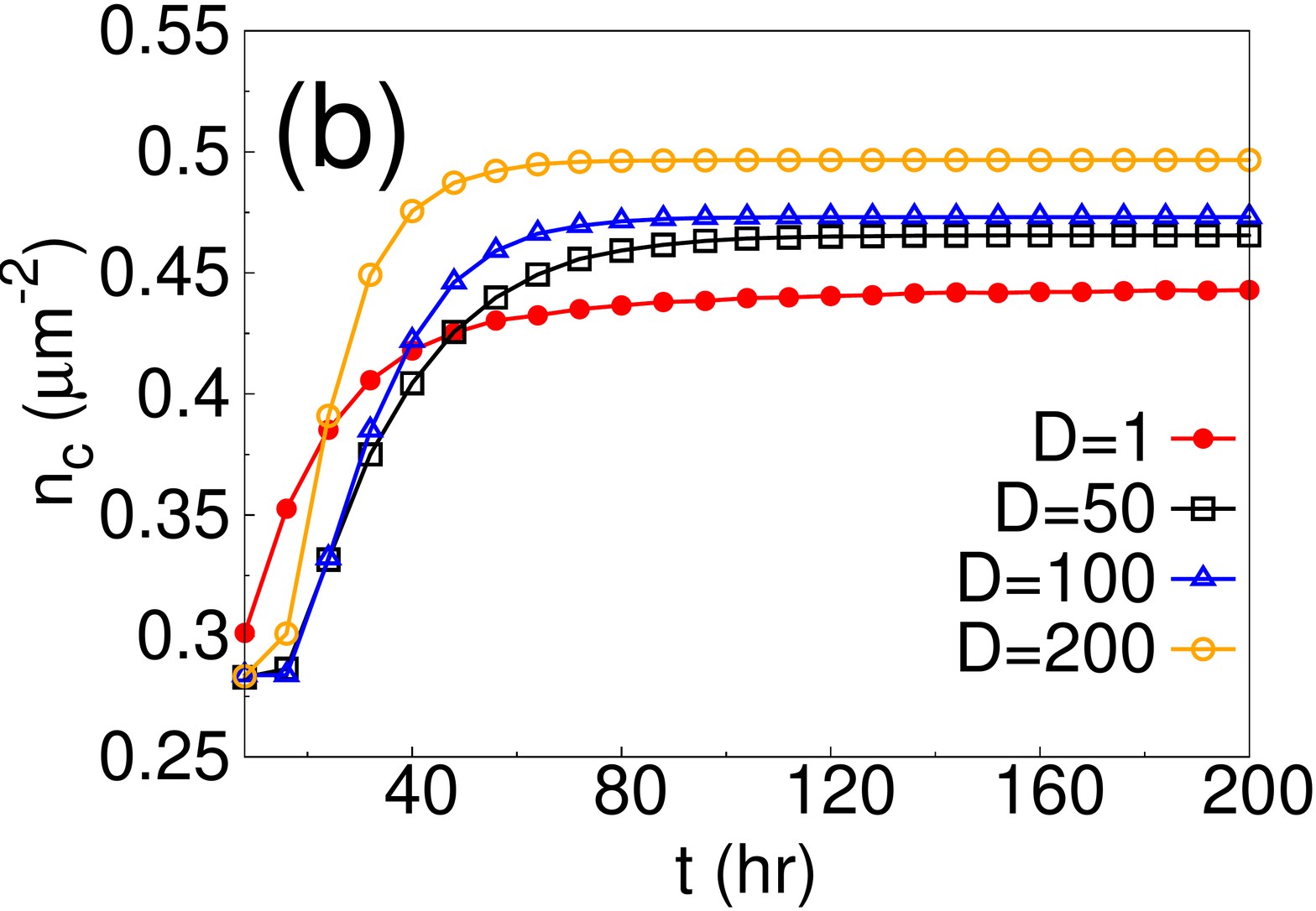}
	\end{subfigure}
	\begin{subfigure}	\centering
		\includegraphics[width=0.475\linewidth]{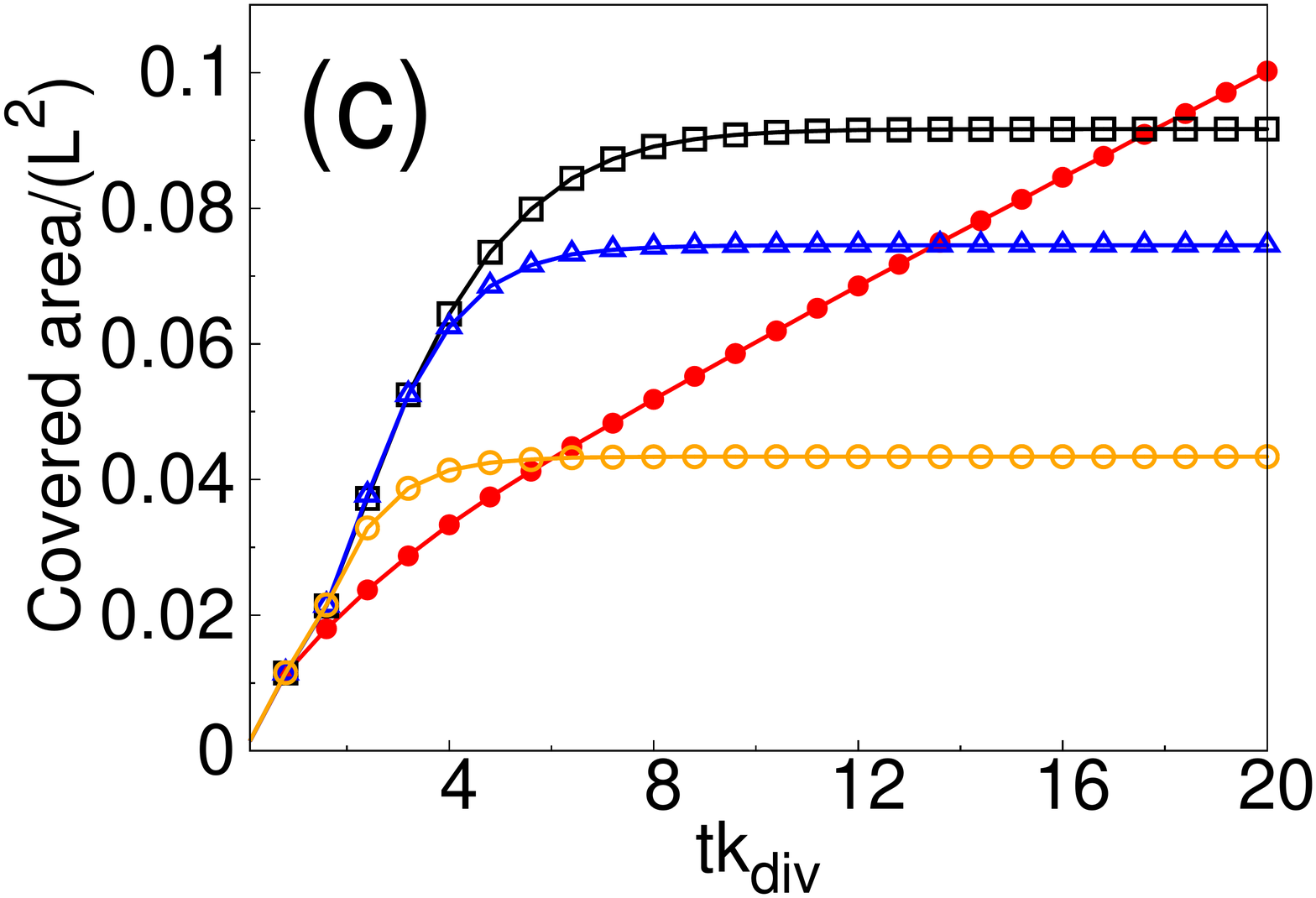}
	\end{subfigure}
	\begin{subfigure}	\centering
		\includegraphics[width=0.475\linewidth]{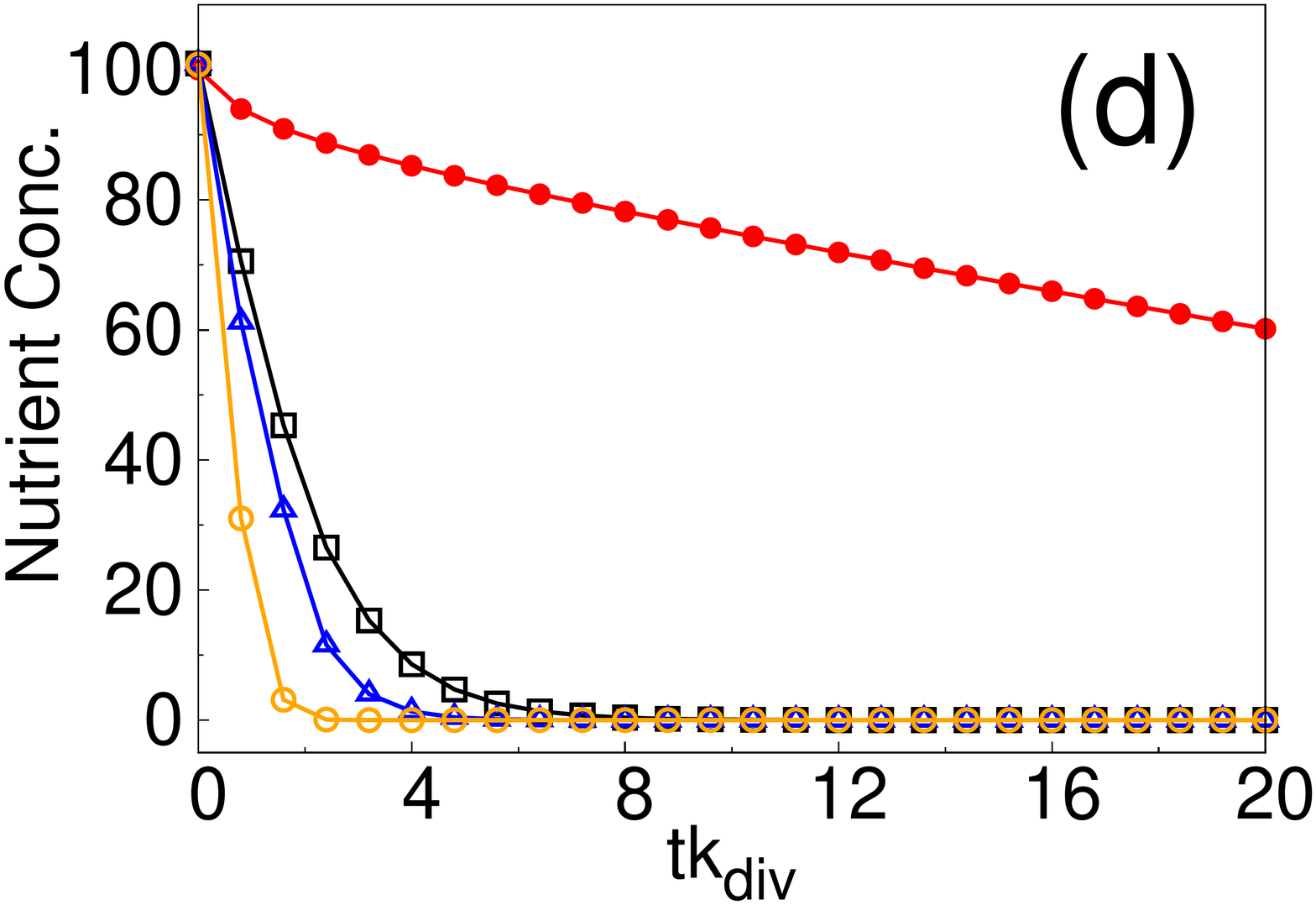}
	\end{subfigure}
	\begin{subfigure}	\centering
		\includegraphics[width=0.475\linewidth]{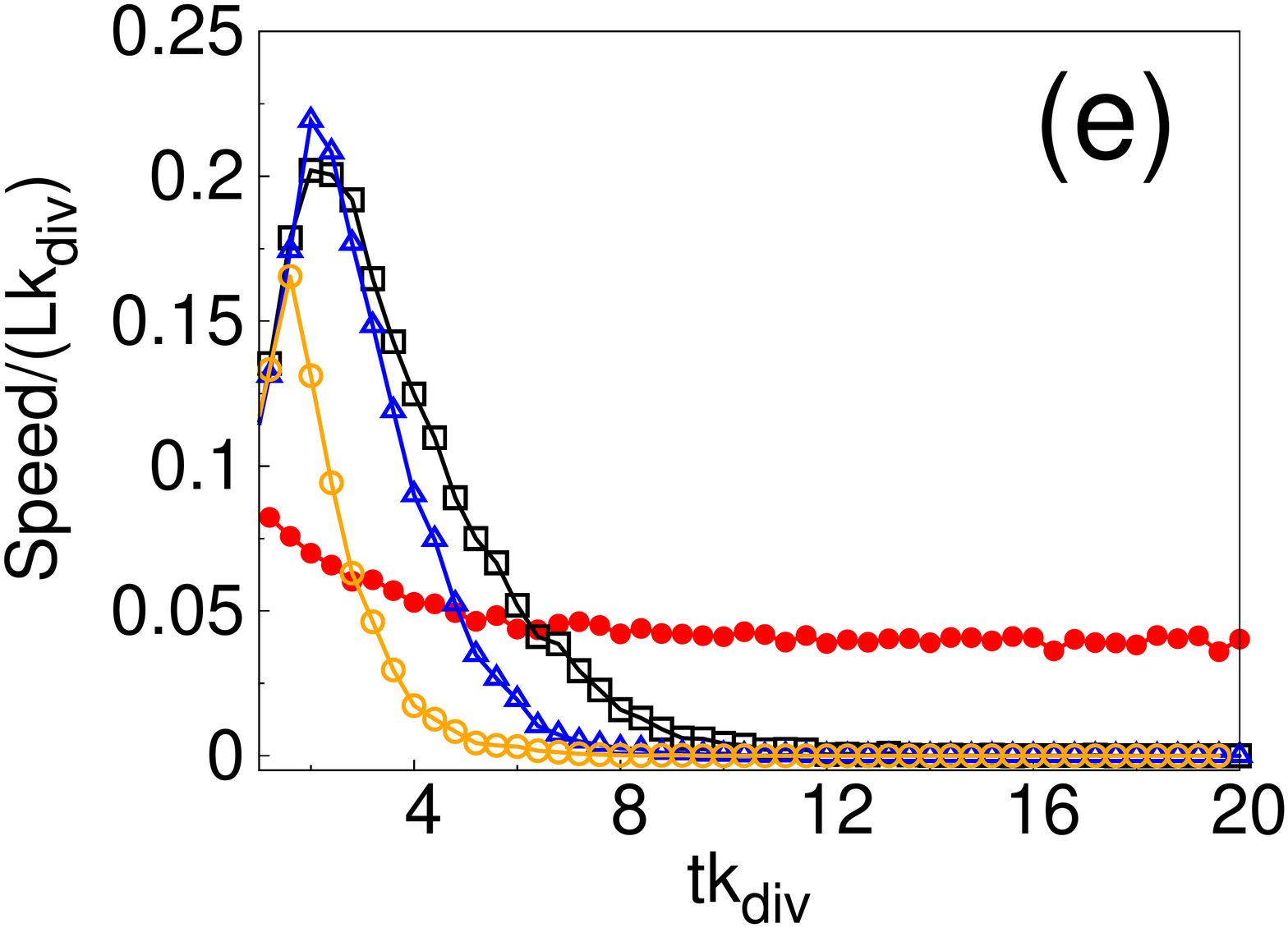}
	\end{subfigure}
	\caption{Bacteria growth dynamics due to the variation of diffusion coefficient of nutrient. (a) Snapshots of colonies from our agent-based simulations for different nutrient diffusivity $D$, \ADD{(b) Time evolution of the  cell-number density $n_c$,} (c) Plot of area covered by bacteria as a function of time, (d) Plot of nutrient concentration with respect to time and (e) Plot of speed of the colonies with respect to time. Each curve in figures(b-e) corresponds to different values of nutrient diffusion coefficients: red curve with filled red circle($D=1$), black curve with empty black square ($D=50$), blue curve with empty blue triangles($D=100$) and orange curve with empty orange circles($D=200$). 
		All the other parameters are chosen to be the same as given in Table-1, except the initial nutrient concentration is $C_0=100$.}
	\label{diffusion_effect}
\end{figure}

To gain further insight on the role of nutrients in colony growth and its morphology, we now vary the value of diffusion coefficient $D$ for a fixed initial nutrient concentration $C_0=100$. We find that the colony morphology changes from branched to smoother fronts as we increase the diffusion coefficient from $D=1$ to $D=200$ [see Fig.~\ref{diffusion_effect}(a)]. For small values of nutrient diffusivity $D=1$ only the cells at the frontier get nutrients and thereby grow and divide. On the other hand, for large values of $D\geq50$, the local nutrients utilized by cells are quickly replenished from the surrounding regions because of high diffusivity. The cells in the interior as well as the front keeps on multiplying filling up densely the entire space until the overall nutrient concentration becomes negligible. 

\ADD{We find that the cell number density $n_c$, where $n_{c}$ = (total number of cells/covered area by the cells) initially increases and reaches a steady-state over long times [see Fig.~4(b)].  The growth in $n_c$ is fastest for higher $D$ as the nutrients are replenished faster. The slight variation $ \sim 5\%$ in the steady state cell-density is attributed to the variation in the bacterial sizes of the population.}  
More intriguingly, the plot of the covered area by the cells versus time [Fig.\ref{diffusion_effect}(c)] shows that for the case of high $D$, as mentioned earlier, the colony spreads exponentially fast until it reaches saturation. On the other hand the 
colony growing under low diffusion coefficient grows slowly but spreads to a much larger area. This is because, as mentioned earlier,  only the cells at the frontier get nutrients to grow and divide and the number of inactive cells increases in the bulk and keeps on growing with time. To further validate our explanation, in Fig.\ref{diffusion_effect}(d) we show that nutrients are rapidly depleted in colonies with high $D$. The rapid consumption of nutrients leads to a dramatic rise in the front speed for  colonies with large $D$  [Fig.\ref{diffusion_effect}(e)] on the other hand, for the colony with $D=1$ the front speed attains a near constant speed because of balance in nutrient consumption by the cells at the frontiers and their reproduction.

\ADD{
	As mentioned earlier, we will now present a continuum model where we incorporate population fluctuations, to further study the
	role of nutrient concentration and population fluctuations in bacterial colony growth.
}
\section{Continuum Model}
\label{sec:fish}
We now consider a nutrient-bacteria (NB) model  in which bacteria consumes nutrient and divides at a rate $\gamma$ per unit biomass while diffusing through space. In the mean-field setting, the bacteria-nutrient dynamics can be described by the diffusive Fisher-Kolmogorov equations \cite{kol37,fis37,gol98}: 
\begin{eqnarray}\label{mean-field}
\partial_t c &=& D \nabla^2 c - \gamma \rho_B c,~\rm{and} \\
\partial_t \rho_B &=&D_B \nabla^2 \rho_B +\gamma\rho_B c.
\end{eqnarray}
Here, $\rho_B(\bm{x},t)$ is the bacterial number density and $c(\bm{x},t)$ is the nutrient concentration at position ${\bm x}$ and time $t$, $D_B$ and $D$ are diffusion coefficient of bacteria and nutrient, \ADD{and total number density over the entire domain i.e. $\rho_T\equiv \int [\rho_B({\bm x},t) + c({\bm x},t)] dx dy/L^2$} remains conserved.  Variants of NB model  ($Eq.~\ref{mean-field}$) but with more complicated reaction and diffusion terms have been used earlier to investigate the transition from type-A to type-B \cite{fis37,gol98,sch16}. However, these mean-field models ignore the role of population fluctuations in the system. However, recent studies have shown that population fluctuations cannot be ignored and are crucial in determining the statistics of growth front \cite{doe03, kor10, nes14}. In particular, Kessler et al. \cite{Kessler1998} using particle based simulation  had indicated that population fluctuations can lead to destabilisation of bacterial colony front spreading by consuming nutrients.

We incorporate population fluctuations in the NB model by adding to it a multiplicative noise term similar to stochastic Fisher-Kolmogorov-Piscunoff-Petrovsky equation \cite{doe03,pig13,pig14}. We show that population fluctuations, inherent to any agent-based model (see Section \ref{sec:ind}), can lead to a transition from type-A to type-B colony morphology.  The stochastic NB (sNB) model that we use, written in terms of the total density $\rho({\bm x},t)$ and the bacterial number density $\rho_B({\bm x},t)$, are  
\begin{eqnarray}\label{stochastic}
\partial_t \rho &=&D_B \nabla^2 \rho_B +D \nabla^2 c, ~\rm{and} \\
\nonumber
\partial_t \rho_B &=&D_B \nabla^2 \rho_B +\gamma\rho_B(\rho - \rho_B) +\mu\sqrt{\rho_B(\rho-\rho_B)}\eta(\bm{x},t), 
\end{eqnarray}
where $\eta(\bm{x},t)$ is a Gaussian white noise  with $\langle \eta({\bm x},t) \rangle=0$, $\langle \eta({\bm x},t) \eta({\bm x}',0) \rangle= {\bm \delta}({\bm x - {\bm x}'})\delta(t)$, $\mu$ controls the noise strength,  and $\langle \rangle$ indicate averaging over noise realizations. 

In the above discussion we have assumed that the motion of the bacteria is independent of nutrient concentration. However in a more realistic case -- similar to the earlier discussed agent-based model -- the motility of the colony might depend upon food as well, wherein scarce food conditions will lead to very less or no movement at all. Following Ref.~\cite{gol98}, we incorporate this effect by replacing bacterial diffusivity term in sNB model (Eq.~\eqref{stochastic}) by a non-linear food dependent diffusivity to get sNBNL model,
\begin{eqnarray}\label{stochastic-nonlinear}
\partial_t \rho &=&D_B \nabla^2 \rho_B +D \nabla^2 c, ~\rm{and}\\
\nonumber
\partial_t \rho_B&=&D_B {\nabla}.(c {\nabla}\rho_B) +\gamma\rho_B(\rho - \rho_B) +\mu\sqrt{\rho_B(\rho-\rho_B)}\eta.
\end{eqnarray}

In what follows,  we present a systematic study of how bacterial front speed and morphology is modified because of nutrient concentration, population fluctuations, bacterial diffusivity, and nutrient diffusivity using direct numerical simulations (DNS) of sNB- and sNBNL-model [Eqs.~\eqref{stochastic} and  \eqref{stochastic-nonlinear}].

\subsection{Numerical Simulations}
We perform simulations for Eq.\eqref{stochastic} and Eq.\eqref{stochastic-nonlinear} in a square domain of  length $L$ and discretize it using $N^2$ collocation points. 
All the spatial derivatives are evaluated using a second order centered finite-difference scheme. For time marching, we use a variant of the operator splitting scheme proposed 
in Refs.~\cite{Dornic2005,Pechenik1999} (see Appendix~\ref{appendix}).  We initialize the bacterial number density as $\rho_B(\bm{x},0) =\frac{1}{2}[1-\tanh\{a(y-b)\}]$ and 
the initial nutrient concentration as $c(\bm{x},0) = C_0 [1-\rho_B(\bm{x},0)]$. The  constants $a$ and $b$ prescribe the width and the position of the colony front and $C_0$ fixes the 
initial nutrient concentration. We impose Neumann Boundary conditions on all sides of the simulation domain for both $\rho_B(\bm{x},t)$ and $c(\bm{x},t)$. Since in the macroscopic 
experiments the number of bacteria that constitute to the colony are large, we fix the strength of population noise to a small value $\mu=5 \times 10^{-2}$ in all our simulations.

\begin{figure*}[ht!]
	\centering
	\includegraphics[width=0.99\linewidth]{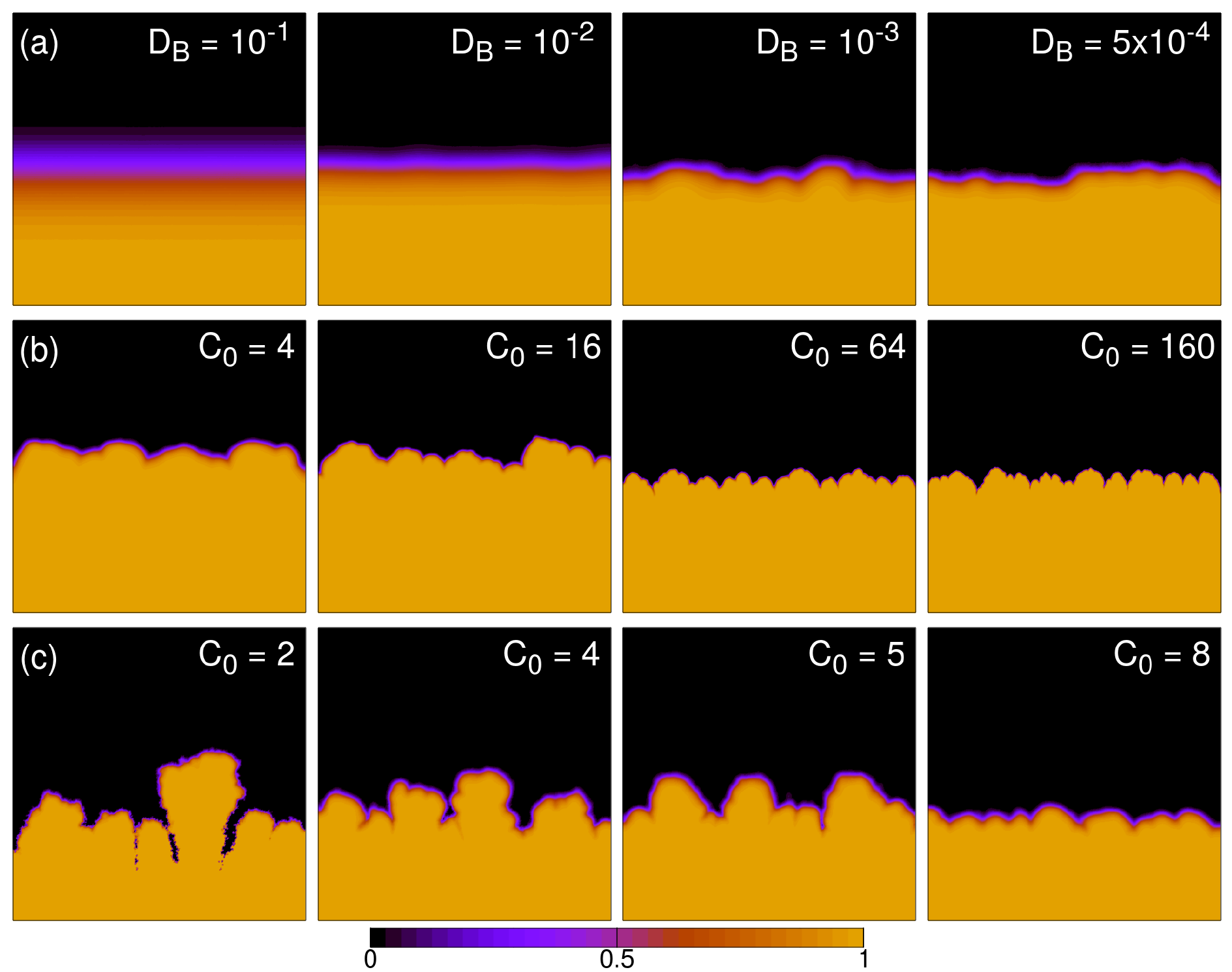}
	\caption{Snapshots from the numerical simulations at different times but comparable colony size for linear diffusion case ($L=10,N=1000$) in rows (a),(b) and non-linear diffusion  ($L=10,N=500$) in row (c). Concentration profiles ($\rho_B/\rho$) are depicted at different times (Black = 0, Yellow = 1) to show the effect of changing diffusion coefficient $D_B$ [$C_0 =1$, row (a)] and nutrient concentration $C_0$ [$D=5\times 10^{-4}$, row (b),(c)] on the spatio-temporal morphology of the growing colony (value in top right corner of each image) keeping rest of the parameters fixed for linear-diffusivity. Parameters are $D=10^{-1}$, $\mu=5 \times 10^{-2},\gamma=1$.}
	\label{snapshot}
\end{figure*}

\subsection{Simulation Results}
The plot in Fig.~\ref{snapshot} shows a representative snapshot highlighting the changes in the front morphologies for different values of initial nutrient concentration $C_0$ and $D_B$. In the following sections, 
we present a systematic study to quantify these morphological patterns.
\subsubsection{Front Speed}
\begin{figure}[ht]
	\begin{subfigure}	\centering
		\includegraphics[width=.49\linewidth]{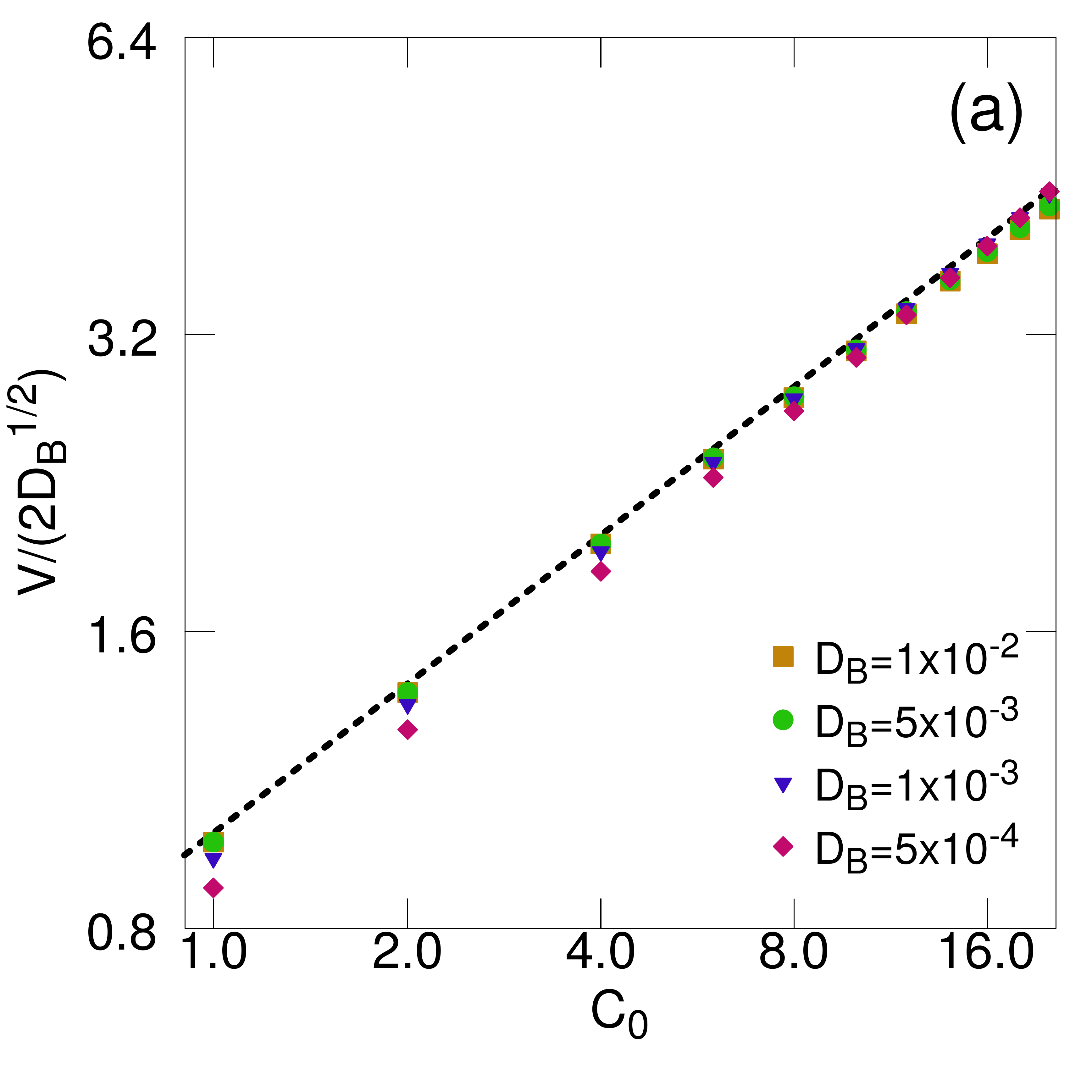}
		\label{vel}
	\end{subfigure}
	\begin{subfigure}	\centering
		\includegraphics[width=.49\linewidth]{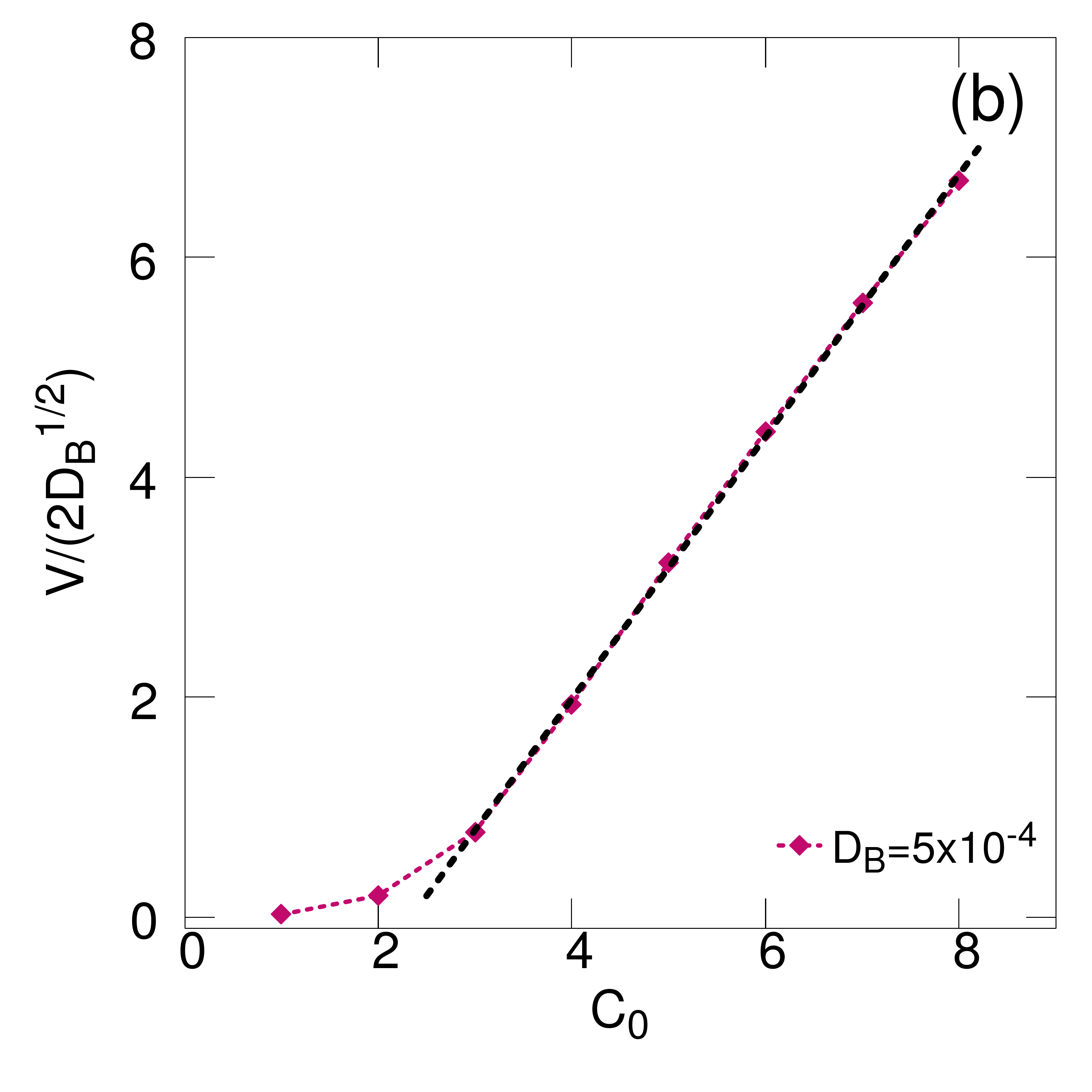}
		\label{vel-nl}
	\end{subfigure}
	\caption{(a) Plot of the front velocity $V$ (scaled with $2\sqrt{D_B}$) versus initial nutrient concentration $C_0$  obtained from DNS of sNB model [Eq.~\eqref{stochastic}] on a log-log scale ($L=32, N=3200,D=1\times10^{-1}, \gamma=1,\mu=5\times10^{-2}$).  The black line show the expected mean-field $V\sim \sqrt{C_0}$ scaling. At lower $C_0$, front velocity is significantly lower than mean-field predictions, which is due to effect of stochastic fluctuations on the system. (b) Plot of the front velocity $V$ (scaled with $2\sqrt{D_B}$) versus initial nutrient concentration $C_0$  obtained from DNS of sNBNL model [Eq.~\eqref{stochastic-nonlinear}] ($L=64, N=4096,D=1\times10^{-1}, \gamma=1,\mu=5\times10^{-2}$), black line shows the linear scaling $V \sim C_0$.}
	\label{velocity}
\end{figure}

An initial linear innoculation of bacteria $\rho_B({\bm x},0)$ spreads outward in Y-direction by consuming nutrients. The speed of this growing colony can be calculated as 
\begin{equation}
V \equiv \frac{d}{dt}\Big<\frac{1}{L}\int_{\Omega}\,\frac{\rho_B(\bm{x},t)}{\rho(\bm{x},t)}\,d\Omega\Big>.
\end{equation}
For Eq.~\eqref{mean-field}, using the marginal stability principle we expect the front speed $V \sim 2\sqrt{\gamma D_B {C_0}}$ \cite{Kessler1998}.  We now investigate how the front propagates for sNB- and sNBNL-model.
\begin{itemize}
	\item {sNB model, Eq.~\eqref{stochastic}} -- In Fig.~\ref{velocity}(a), we plot front speed versus concentration for fixed  $D_B=5 \times 10^{-4}, D=10^{-1}, \mu=5\time 10^{-2},$ and $\gamma=1$. Although the 
	colony morphology changes on changing  $C_0$ and $D_B$, we find that the mean-field prediction of front-velocity is in excellent prediction with numerics i.e., $V\sim \sqrt{C_0}$. 
	\item {sNBNL model, Eq.~\eqref{stochastic-nonlinear}} -- The plot in Fig.~\ref{velocity}(b) shows that the front speed scales linearly with the initial nutrient concentration for $C_0 \geq 3$. At leading order,  we can approximate the nonlinear diffusion term $D_B \nabla \cdot c \nabla \rho_B$ as $D_B C_0 \nabla^2 \rho_B$. Thus by making an analogy with Eq.~\ref{stochastic} we expect $V\sim C_0$.  
\end{itemize}
\subsubsection{Morphological behavior}
Our numerical simulations show that population fluctuations give rise to diffusive instabilities in the propagating front \cite{Kessler1998} which lead to different kinds of morphological behavior  that are absent in mean-field equations (NB model). Fig.\ref{snapshot}(a) shows the colony morphology with varying bacterial diffusivity, where the front width decreases with decreasing $D_B$, with rough front appearing at smaller $D_B$.

\begin{figure}[ht!]
	\begin{subfigure}
		\centering
		\includegraphics[width=.49\linewidth]{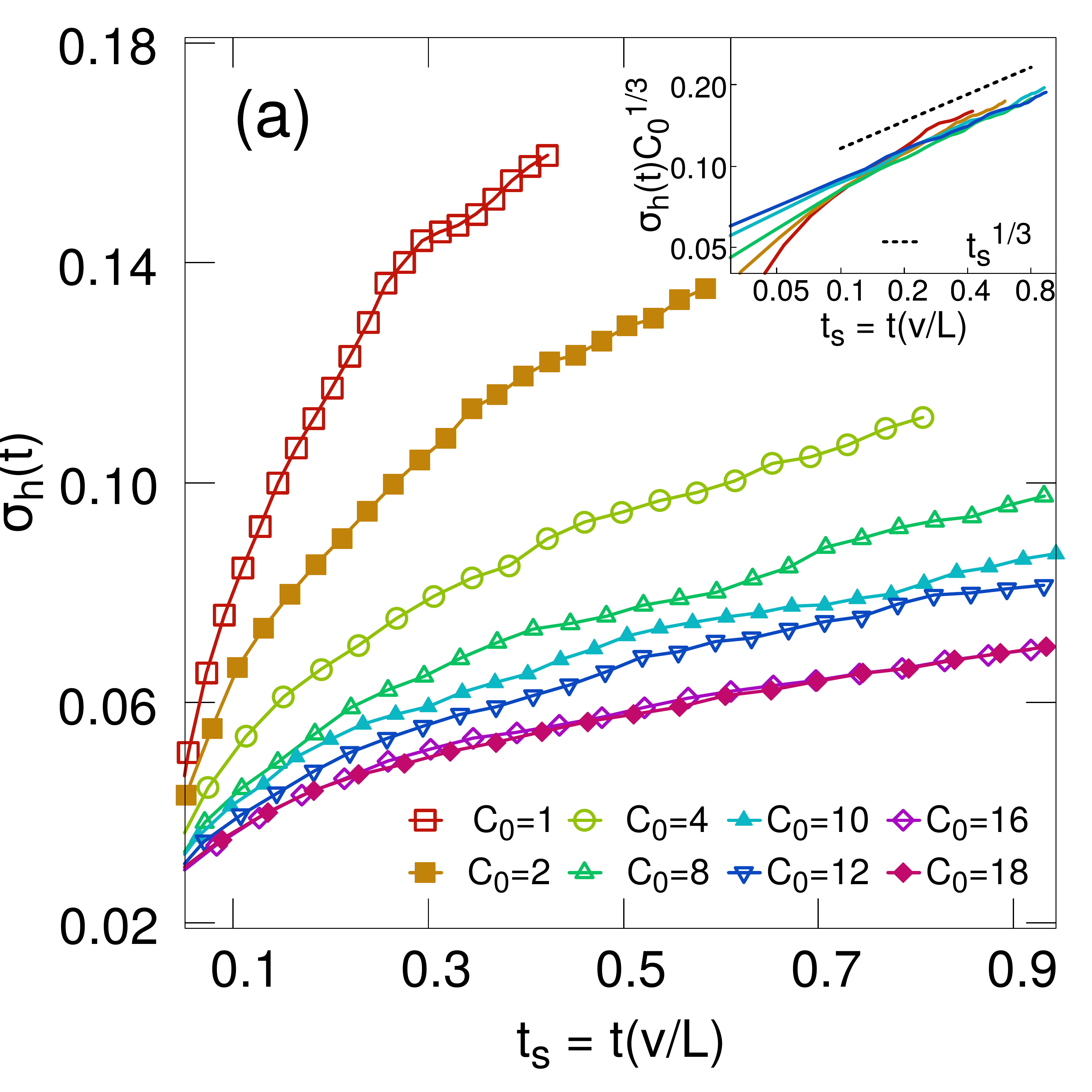}
		\label{stats1}
	\end{subfigure}
	\begin{subfigure}
		\centering
		\includegraphics[width=.49\linewidth]{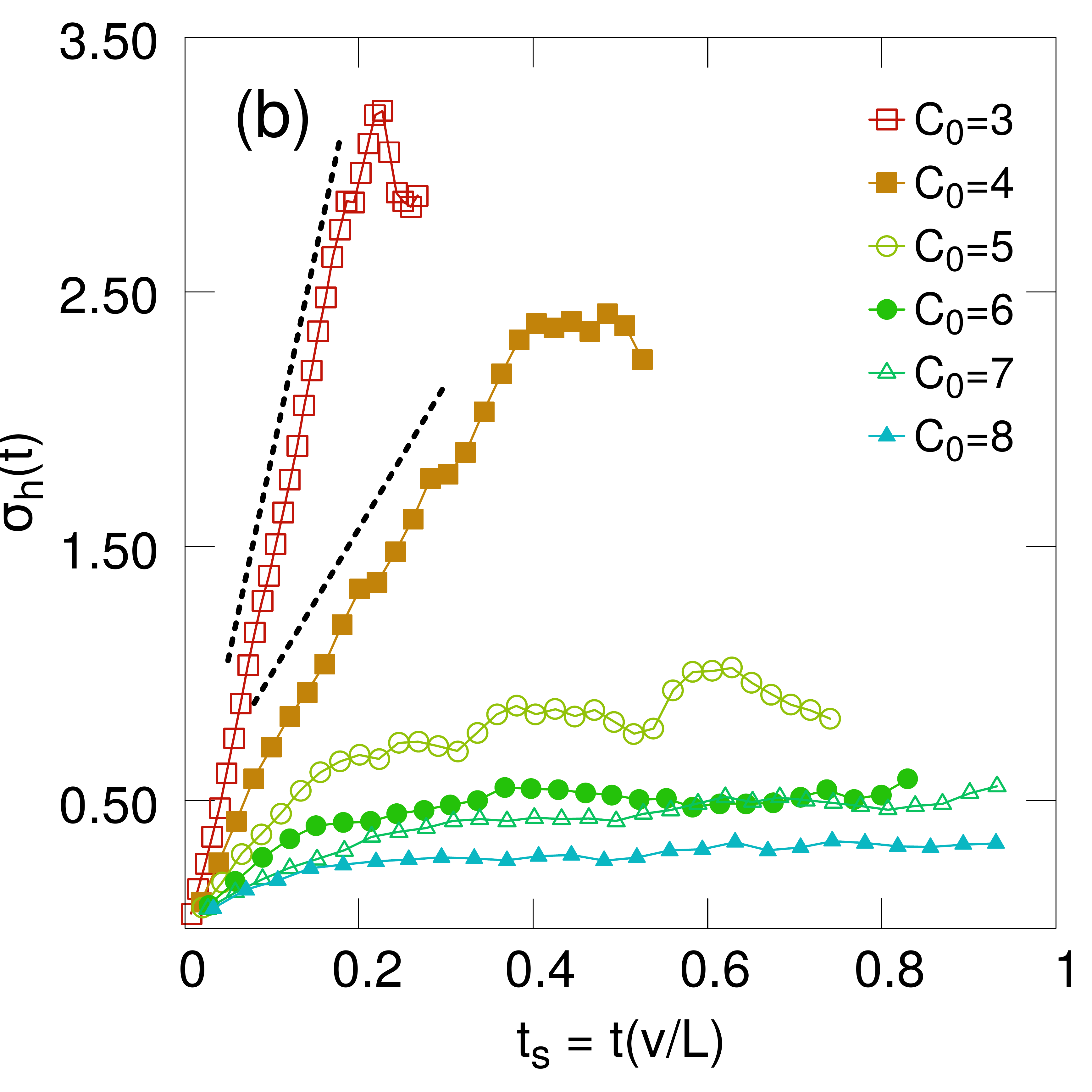}
		\label{stats2}
	\end{subfigure}
	\caption{Roughness $\sigma_h(t)$ versus time (scaled with $V/L$) for different nutrient concentration, $C_0$, ${(a)}$ for sNB model ($L=32,N=3200$), the inset in $(a)$, shows the plot of $\sigma_h(t) C_0^{1/3}$ vs time(scaled), showing data collapse over $t^{1/3}$ line, which is in agreement with the exponent found for stochastic Fisher equation in nutrient rich conditions \cite{nes14}. $(b)$ For the sNBNL model ($L=64,N=4096$), for small values of $C_0$, $\sigma_h(t)\sim t$ (black dashed line) and  saturates for higher values of $C_0$ ($D_B=5\times10^{-4},D=10^{-1},\gamma=1,\mu=5\times10^{-2}$). This is consistent with our findings in for the agent based model in Section~\ref{sec:ind}. Sudden drops in $\sigma_h(t)$ are due to merging of different branches. Data for sNBNL model is from one ensemble only.}
	\label{stats} 
\end{figure}

The plot in Fig.~\ref{snapshot}(b) shows the role of nutrient concentration on growth dynamics for sNB model. We find that the front undulations decrease on increasing $C_0$. We quantify the front undulations by plotting $\sigma_h(t) = {\Big<\overline{[h(x,t)-\overline{h}]^2}\Big>}^{1/2}$ \cite{Farrell2013,Bonachela2011,Barabasi_B95} where $h(x,t)$ is the height of the front, the bar means spatial average in $x$ direction and angular brackets denote ensemble average. As expected, we find that $\sigma_h$ increases with decreasing $C_0$ [see Fig.~\ref{stats}]. For the sNB model,  similar to Ref.~\cite{Nesic}, we find that $\sigma_h(t)\sim t^{1/3}$. On the other hand in the sNBNL model, similar to the agent-based model, the dynamics of the front structure dramatically alters on varying the nutrient concentration. Small values of $C_0$ gives rise to more prominent finger like patterns and $\sigma_h(t)\sim t$ [see Fig.~\ref{snapshot}(c)]. On further increasing $C_0$, finger like growth transitions into a smooth and compact front [see Fig. ~\ref{stats}(b)].

Our results are in qualitative agreement with agent-based simulations. Our results show that the population noise along with non-linear diffusion (sNBNL model) is sufficient to show the morphological transition from finger/branched fronts to smooth fronts.

\section{Conclusions}
We studied the role of nutrients \ADD{and population fluctuations} on the spreading of bacterial colony on a hard agar plate using both agent based and continuum simulations. We find a qualitative agreement between the two methodologies. The main conclusions of our study are:
\begin{enumerate}[label={(\roman*)}]
	\item Initial nutrient concentration has profound effect on colony growth leading to morphological changes. A systematic change of initial nutrient concentrations from lower to higher values causes transition of the colony periphery leading to the formation of finger-like to branched-like structure to smoother front,
	\item Roughness of the colony front decreases with increase in initial nutrient concentration. In particular, for small values of $C_0$, both agent based simulations and sNBNL model shows that $\sigma_h \propto t$. 
	\item Front speed of the colony increases as a function of initial nutrient concentration and follows the mean field prediction for the sNB model $V\sim\sqrt{C_0}$~\cite{Kessler1998} 
	and sNBNL model $V\sim C_0$. These predictions are in qualitative agreement with the agent based model.
	\item Our continuum simulations indicate that population fluctuations, inherent to agent based models, play crucial role in the formation of various morphological patterns.
\end{enumerate}
\ADD{Although our present model only considers bacterial growth in a monolayer on surface, there is a definite scope to extend our model in three-dimensions to study the growth dynamics of bacteria forming biofilm-like structures. Bacteria growth and development in three-dimensions might lead to complex morphologies as an outcome of interactions of bacteria with surface and extracellular matrix~\cite{Asally}. Moreover, it would also be interesting to investigate the spatiotemporal dynamics of coexisting species using the ideas we present here.}

\vspace{0.5cm}
\textbf{Acknowledgement:}
We thank Jagannath Mondal for discussions. This work is partially supported by DST-INSPIRE Faculty Award[Pushpita Ghosh/DST/INSPIRE/04/2015/002495].

\section*{Appendix}
\subsection{Numerical Integration of Stochastic Part}
\label{appendix}
We follow the algorithm suggested in Ref.\cite{Dornic2005} to numerically integrate Eq.\eqref{stochastic} and Eq.\eqref{stochastic-nonlinear}.
We use operator-splitting scheme to first solve the stochastic part i.e.
\begin{eqnarray}\label{full-langevin}
\frac{d\rho}{dt} = \sigma \sqrt{\rho(1-\rho)}\eta(x,t)
\end{eqnarray}
Here $\eta(x,t)$ is a random normal deviate.
We approximate $\rho(1-\rho)$ as $\Theta(1/2-\rho) \times \sqrt{\rho}$ + ($\rho \leftrightarrow 1-\rho$) \cite{Dornic2005}.
The effective equation to be solved then is
\begin{eqnarray}\label{approx-langevin}
\frac{d\rho}{dt} = \sigma \sqrt{\rho}\eta(x,t)
\end{eqnarray}
for which the associated Fokker-Planck equation and it's solution are
\begin{eqnarray}\label{fokker-planck}
\partial_t P(\rho,t) &=& \frac{\sigma^2}{2}\partial^2_\rho [\rho P(\rho,t)] \\
P(\rho,t) &=& \delta(\rho)e^{-2\rho_o/\sigma^2 t} + \frac{2e^{-2(\rho_o+\rho)/\sigma^2 t}}{\sigma^2 t} \sqrt{\frac{\rho_o}{\rho}}I_1\Bigg(\frac{4\sqrt{\rho_o\rho}}{\sigma^2 t}\Bigg)
\end{eqnarray}
Thus we have $\rho(t+dt) = \rho^*$ where $\rho^*$ is random number from the distribution $ \rho^* = \textrm{Gamma}[\textrm{Poisson}[\lambda \rho(t))]]/\lambda$ where $\lambda = \frac{2}{\sigma^2 dt}$.
The deterministic part is then solved using Euler's Method for Eq.\eqref{stochastic} and Adam-Bashford scheme for Eq.\eqref{stochastic-nonlinear}.

\end{document}